\documentstyle[prl,aps,epsfig,preprint]{revtex}

\newcommand{\dd}{\partial}

\newlength{\textwidthm}
\begin{document}

\draft

\preprint{}

\title{Nonlinear Optics and Quantum Entanglement of Ultra-Slow Single
Photons    }
\author{M.~D.~Lukin$^1$ and  A.~Imamo{\u g}lu$^2$}
\address{
	$^1$ ITAMP, Harvard-Smithsonian Center for Astrophysics,
		Cambridge, Massachusetts~~02138 \\
	$^2$ Dept. of Electrical and Comp. Engineering, and Dept. of Physics,
University of California, Santa Barbara, CA 93106
}
\date{\today}
\maketitle

\begin{abstract}

Two light pulses propagating with ultra-slow group velocities in a coherently
prepared atomic gas exhibit dissipation-free  nonlinear coupling of an
unprecedented strength. This enables a single-photon pulse to
coherently control or manipulate the quantum state of the other. Processes of
this kind result in  generation of entangled states of  radiation field and
open up new prospectives for  quantum information  processing.

\end{abstract}

\pacs{PACS numbers 42.50.-p, 42.65.-k, 42.50.Dv,03.67.-a}

\newpage

It has been known for more than thirty years that light fields or photons
can interact with each other in atomic media much like massive
particles do \cite{bloom}. However, the strength of the interaction of two
single light quanta is typically extremely weak. As
a  result, conventional nonlinear optics is feasible only when  powerful
laser beams, containing a large number of photons, interact in nonlinear
materials.

This Letter describes a method that allows for two slow light pulses
\cite{hau99,kash99,budker99} of tiny energies to
interact in  a resonant  ensemble of atoms.
When Electromagnetically Induced Transparency (EIT)
\cite{scullybook,harrisrev} is established for {\it
both of these pulses}, they will propagate with ultra-slow
but equal group
velocities, and a very efficient nonlinear interaction between
them will take place \cite{atac96}. This interaction can be maintained for a
very long time  without dissipation, resulting in a new regime of
quantum nonlinear optics. Specifically we describe here a
scheme in which a traveling light pulse with an energy
corresponding to that of a ``single photon'' ($\hbar \nu$) can modify the
refractive index of a second traveling pulse such that the latter
experiences a  non-linear phase shift on the order of $\pi$. This process
allows one to create strongly correlated (i.e. entangled) states
\cite{scullybook}
of interacting photons and to generate macroscopic
quantum superpositions such as Schr{\"o}dinger cat states \cite{cats}.
The technique described here opens up interesting prospects for coherent
processing of quantum
information \cite{Ekert96,qc,lloyd95} in applications such as quantum
computation and quantum communication.

It has been widely accepted that the use of ultra-high finesse
micro-cavities is essential for having strong interactions between single
photons \cite{kimblerev,atac97,walls}. In the present contribution
a radically different approach  is taken,
 in that the properties of an optical material are ``designed''  to
achieve an unprecedented strength of nonlinearity without the need for a
cavity. Earlier work has already  demonstrated that such a coherently
controlled dense atomic ensemble displays remarkable phenomena
\cite{scullybook,harrisrev}, which are of great interest for nonlinear optics
at a low light level
\cite{steve98,lukin99,steve99,frans}.

We consider a process in which a pair of very weak
optical  fields $E_1,E_2$ interact resonantly with an ensemble of $N_1$
multi-state atoms (species A) as depicted in Fig.1. These atoms
are coherently driven by a classical (i.e. many-photon) light
field with a Rabi frequency $\Omega_1$
tuned to resonance with an atomic transition $|c_1 \rangle\rightarrow
|a_1\rangle$.  We assume that the frequency of
one of the weak optical fields ($E_1$) is tuned to resonance with transition
$|b_1 \rangle \rightarrow |a_1 \rangle$. Quantum interference induced by the
driving field results in a sharp transmission resonance in the spectrum of
the weak field $E_1$ (Fig. 2). The second weak field
($E_2$) couples another optically allowed
transition  $|c_1 \rangle \rightarrow |d_1 \rangle$ with a single photon
detuning $\Delta$. When $E_2$ is absent a properly tuned
$E_1$ propagates without
loss or refraction, but its group velocity is substantially reduced:
this is the essence of EIT.
The off-resonant field $E_2$ induces a Stark shift of the state
$|c_1\rangle$ and
therefore  modifies the refractive index of the first field.  Resulting
phenomenon  corresponds to cross-phase modulation of two weak light waves
\cite{atac96}: since the dispersion of refractive index is very steep
(Fig.2), small Stark shifts result in a large index change.

The key idea of the present work is to arrange the conditions such
that the two weak light pulses  can interact in transparent, nonlinear
media for a very long
time. This is the case  when pulses propagate with equal, slow group
velocities.  Photon interaction will then
cause large nonlinear phase shifts even if pulses of
ultra-small energies are involved. A possible technique to achieve equal
group velocities for two weak pulses is explained in Fig.1: here
$N_2$ $\Lambda$-type atoms of
a second kind (B) are added to the interaction region, and
the frequency of the field $E_2$ is chosen to be on resonance
with the optical transition
$|b_2\rangle \rightarrow |a_2\rangle$. Application of a second
driving classical field $\Omega_2$ of appropriate intensity will result
in induced transparency and properly reduced  group velocity for the
field $E_2$. The mixture of atoms required to achieve such group velocity
matching can be designed by using different
isotopes of alkali atoms, and by applying an appropriate magnetic field to
create the desired level splittings.

We describe quantum radiation by multi-mode field operators
\cite{scullybook}: $
{\hat E}_{1,2} (z,t) = \sqrt{2\pi c/ L} \sum_k {\hat a}_{k1,2}(t)
e^{-i k z}$, where ${\hat a}_{k i}$ is the annihilation operator
corresponding to the field mode with wave-vector $k$.
$L$ is the quantization length and $c$ is the speed of light in vacuum.
In the present
approach we choose to restrict the quantization to a continuum of modes with
wave-vectors ranging from $k_0 - \Delta \omega /2c$ to $k_0 + \Delta \omega
/2c$, which in turn gives equal space-time commutation  relations
$[{\hat E_i}(z,t),{\hat E_j}^+(z,t)] = \Delta\omega \delta_{ij}$,
that depend explicitly on the bandwidth of the system
$\Delta \omega$. This bandwidth can be determined, e.g., by
the detection process. As we will discuss shortly, the spectral width of the
transparency window also gives a natural limitation on the maximal
bandwidth.

The dynamics of the atomic media is described by Heisenberg
equations for atomic polarizations and coherences. The finite bandwidth
of  quantized fields allows us to apply
adiabatic elimination of the atomic  degrees of freedom.
By  disregarding time derivatives of third and higher
order,  we arrive at the following  evolution equations for the field
operators:
\begin{eqnarray}
\label{slowly-varying1}
\left(\frac{\dd}{\dd z} + \frac{1}{v_{g}}
\frac{\dd}{\dd t}
\right){\hat E}_1 = -\kappa {\hat E}_1  &+&
\beta \frac{\dd^2}{\dd t^2} {\hat E}_1  \nonumber \\
&+& i \eta {\hat E}_2^+ {\hat
E}_2
{\hat E}_1 + {\hat F}_1 \\
\label{slowly-varying2}
\left(\frac{\dd}{\dd z} + \frac{1}{v_{g}}
\frac{\dd}{\dd t}
\right) {\hat E}_2 = -\kappa {\hat E}_2 &+&
\beta \frac{\dd^2}{\dd t^2} {\hat E}_2 \nonumber \\
&+&i \eta {\hat E}_1^+
{\hat E}_1 {\hat E}_2 + {\hat F}_2 .
\end{eqnarray}
The adiabatic expansion procedure used in derivation of
Eqs.~(\ref{slowly-varying1},\ref{slowly-varying2}) is analogous to that of
Refs.\cite{steve98,steve99} for classical fields.  Here, $v_{g} =
c/(1+n_{g})$ is the group velocity for the corresponding pulse, $n_g = g^2
N/|\Omega|^2$, and $\kappa$ being the rate   of residual single-photon loss
$\kappa = n_g \gamma_{bc}/c$.
${\hat F}_1$ and ${\hat F}_2$ are delta-correlated noise
operators associated with dissipation. $\eta = [n_g g^2/
(i \gamma_{cd} + \Delta)] l/(2\pi c^2)$ is the rate of nonlinear interaction
between pulses and
$\beta = n_g \gamma_{ab}/(|\Omega|^2 c)$. $g =
\sqrt{\gamma_{ab} \sigma c/(2Al)}$ is a 
normalized atom-field coupling constant, 
$A$ is the cross-sectional area of the quantized fields,
$\sigma$ is resonant absorption cross-section and
$l$ is the length of the interaction region.
$\gamma_{ij}$ are linewidths of $|i\rangle\rightarrow |j\rangle$ atomic
transitions. We assumed that  all interacting atoms have 
identical linewidths
and coupling constants, and that the number of atoms are
equal ($N_1=N_2 = N$). To obtain equal group velocities, identical
Rabi-frequencies for the driving fields were chosen ($\Omega_1 = \Omega_2 =
\Omega$).

Eqs.~(\ref{slowly-varying1},\ref{slowly-varying2}) together with the
commutation relation define a quantum field theory of interacting pulses in a
coherently prepared medium. In cases when losses can be neglected
this field theory is analogous to that of a weakly
interacting Bose-gas, but with an {\it imaginary} effective mass term.  This
imaginary mass term results from the finite bandwidth of the transparency
window and causes a spreading of the input pulses (Fig.2).

We first consider the classical limit of this theory by replacing
operators ${\hat E_{1,2}}$ with their expectation values
$E_{1,2}$. In the ideal case
$\Delta$ can be chosen to be large and hence $\eta $ is  purely real.
When attenuation and pulse spreading  are small enough to be neglected, Eqs.
(\ref{slowly-varying1},\ref{slowly-varying2}) can be  easily solved by
$E_{1,2}(z,t) = E_{1,2}(0,t') \exp(i\eta |E_{2,1}(0,t')|^2 z)$, which
corresponds to cross-phase modulation of the weak beams. Note that
this result is expressed in terms of retarded time $t'= t- z/v_g$.
Magnitude of the nonlinear phase shift at the peak of the pulse
can be expressed in terms of energy $E$ and duration $T$ of a Gaussian pulse:
\begin{eqnarray}
{\rm phase \; shift} = \sqrt{{\rm ln}(2) \over 4 \pi}
{\gamma_{cd} \over \Delta} {\sigma \over A}
 {E \over \hbar \nu}  {\tau_g \over T}.
\label{phase1}
\end{eqnarray}
Here nonlinearity
$\eta$ is expressed in terms of group delay
$\tau_g = n_g z/c$. In the limit when $v_g\ll c$ the
later  corresponds to the  interaction time of the pulse with the medium.

It is apparent that large phase shifts exceeding $\pi$ at energies
corresponding to a fraction of $\hbar \nu$ appear to be possible when
the interaction time exceeds the pulse duration of a tightly focused
($\sigma \sim A$) laser beam.  Before proceeding it is
important to re-examine the assumptions resulting in this striking
conclusion.

First, we note that a stringent limitation
imposed by the difference of group velocities of interacting pulses
\cite{steve99} is absent in our case since  the interaction is designed
such that group velocities are equal. Second, even in the absence of
various absorption mechanisms,  nonlinear phase
shifts are limited by the bandwidth of the transparency  window which
decreases with propagation distance \cite{lukin97}
$\Delta \omega_{max} = (\beta z)^{-1/2}$. After a sufficiently 
long propagation this results in spreading of the pulse 
(see Fig.2). In order to
avoid losses due to spreading, $\tau_g/T$ should be smaller than
\begin{eqnarray}
\tau_g \Delta\omega_{max} = \sqrt{{\Omega^2 \tau_g\over \gamma_{ab}}} =
\sqrt{{\sigma N\over 2A}}.
\end{eqnarray}
The quantity under square root corresponds to an optical depth of the medium
\cite{lukin97}. Hence it is essential to operate in  
optically dense gas. Finally we  note that the interaction time
($\tau_g$) is itself limited due to absorption associated with
decoherence of the ``dark state'' \cite{harrisrev}; i.e.  $\tau_g <
\gamma_{bc}^{-1}$ must be satisfied.  This implies that EIT
resonances should be strongly saturated, i.e.  $\Omega^2/
\gamma_{ab}\gamma_{bc}
\ge N \sigma/A \gg 1$, in order for nonlinear phase shifts to be large.

We now consider quantum dynamics of the nonlinear interaction.  When
absorption  is  negligible and the bandwidth $\delta$ of the pulses satisfy
$\delta < \Delta \omega < \Delta \omega_{max}$, quantum
Eqs.~(\ref{slowly-varying1},\ref{slowly-varying2}) can be solved  by
\begin{eqnarray}
{\hat E}_{1,2} (z,t) = {\hat E}_{1,2} (t') \; \exp \left[ i \;
\eta
\; {\hat E}^+_{2,1}(t'){\hat E}_{2,1}(t') l \right],
\end{eqnarray}
where ${\hat E}_i(t)$ are Heisenberg operators describing the
input fields at $z = 0$. We first analyze the evolution of most ``classical''
of all possible input wave packets. These are multi-mode coherent
states \cite{scullybook} $|\alpha_1, \alpha_2 \rangle = \prod_{k,j}
|\alpha_{1}^k \rangle\times|\alpha_{2}^j \rangle$, which
are eigenstates of input operators ${\hat E}_{i}(t)$ with
eigenvalues (at $z=0$) $\alpha_{i} (t) = \sqrt{2 \pi c/L} \sum_k \alpha^k_{i}
e^{ikct}$. After  propagation
through the nonlinear dispersive medium, the following expectation
values of the fields can be measured:
\begin{eqnarray}
\langle  {\hat E}_{1,2}(z,&&t) \rangle 
= \alpha_{1,2} (t') \times \; \nonumber \\
&&\exp\left[ [-2{\rm sin}^2(\Phi/2)   + i \;  {\rm
sin}(\Phi)] \; {|\alpha_{2,1} (t')|^2 \over \Delta \omega}
  \right],
\label{al1}
\end{eqnarray}
where quantum phase shift $\Phi = [1/(4 \pi)
\sigma/A \gamma_{cd}/\Delta ] \Delta \omega \tau_g$. These solutions have
the same form as
those obtained in Ref.~\cite{sanders92} for single-mode fields and generalize
that earlier result to the multi-mode case that is appropriate for
the traveling-wave geometry.

Eq.~(\ref{al1}) reproduces the classical  result
{\it only} when $\Phi \rightarrow 0$; for sufficiently large
interaction times the quantum dynamics of wave packets deviates
substantially  from the classical case. In particular, both phases
and amplitudes given by Eq. (\ref{al1}) exhibit periodic collapses and
revivals. The origin of this behavior can be understood by noting
that each component of input coherent states $|\alpha_{1}^k\rangle
|\alpha_{2}^j\rangle$ is itself a coherent superposition of many Fock
components $|n_1^k m_2^j\rangle$. During nonlinear interaction each
of these components acquires a different phase change
resulting, for sufficiently large $\Phi$, in quantum dephasing
of the original coherent states. However
when $\Phi$ reaches multiples of $2\pi$, all components ``re-phase'' such
that original coherent state is reproduced. At intermediate values of
interaction times, re-phasing to other macroscopic states can occur. For
example, when $\Phi = \pi$ the output state of two fields
can be verified to be: $|\psi \rangle = {1 \over {2}} \left(
|\alpha_1,\alpha_2\rangle + |-\alpha_1,\alpha_2\rangle +
|\alpha_1,-\alpha_2\rangle -|-\alpha_1,-\alpha_2\rangle \right)$. 
This is an
entangled superposition of  macroscopically distinguishable  states.
Superpositions of this kind have no  classical counter-part, and correspond
to Schr{\"o}dinger cat-like states \cite{cats}.

Consider now a different type of input quantum state corresponding to
multi-mode single-photon wave-packets: $|1_i\rangle =
\sum_k \xi_k {\hat a}_{k,i}^{\dagger} |0\rangle$
where Fourier amplitudes $\xi_k$ are normalized such that
$\sum_k |\xi_k|^2 =1$.
In free space
these single-photon states represent traveling waves initially ($t=0$)
localized  around $z = 0$. The dynamics of individual wavepackets is  fully
described by  a single photon  ``wave-function''  ${\tilde
\Psi}_{i}(t,z) =
\langle 0  | {\hat E}_i(t) |1_i\rangle$. Correlations of these wavepackets
emerging  due to  photon-photon interaction are described, in turn,
by correlation amplitudes \cite{scullybook}
${\tilde \Psi}_{12}(t_1,z_1;t_2,z_2) = \langle 0
| {\hat E}_1(t_1,z_1) {\hat E}_2(t_2,z_2) |1_11_2\rangle$.  After
propagation though the atomic cell the correlation amplitude is given by:
\begin{eqnarray}
{\tilde \Psi}_{12}(t_1'',z_1&;&t_2'',z_2) = 
{\tilde \Psi}_1(0,t''_1) {\tilde
\Psi}_2(0,t''_2) \nonumber \\
&\times& \left[ 1 
+  {\sl sinc}[\Delta \omega (t''_1-t''_2)/2]
(e^{i\Phi} -1) \right],
\label{twop}
\end{eqnarray}
where $t''_i = t_i-l/v_g - (z_i-l)/c$.
Eq.~(\ref{twop}) indicates that the nonlinear
interaction alters the mode structure of the pulse in addition to generating
a phase shift on existing components. This phenomenon is related to the
classical effect of pulse broadening due to cross-phase modulation.
It is important however that  equal-time (``coincidence'') correlations
$(t''_1-t''_2 )\Delta \omega \ll 1$ indicate that a pair of single photons
acquire a phase shift $\Phi$ as a result of the nonlinear interaction. The
magnitude of $\Phi$ can easily exceed $\pi$, when $\tau_g
\Delta\omega_{max} \gg
1$.

These large nonlinear phase shifts  can be used to create quantum
entanglement.  For example, if the photons are initially in coherent
superpositions of two  states (for example the two
polarization states $|1_{\pm}
\rangle$), and only one of these states ($|1_{+} \rangle$) is subject to the
strong nonlinear interaction, the  resulting state cannot be factorized into
a product state of  individual modes \cite{milburn89}. Operations of
this kind form the  essence of  quantum information
processing \cite{qc}.  The example considered above indicates, however,
that the present approach to quantum entanglement
differs conceptually from the techniques discussed previously. In particular,
usual approaches to quantum processing are
based on systems whose Hilbert space can be restricted to a 2 dimensional
subspace (qubit). In contrast,  the present technique involves multi-mode,
traveling wave excitations where the 2-dimensional Hilbert space of the
qubit (i.e. polarization) and the external degrees of freedom of the
photon field become coupled. It is interesting to consider
how various concepts of
quantum information theory can be applied in the context of our approach.

It is clear that the feasibility of
quantum entanglement depends upon the large delay-bandwidth products
$\Delta \omega_{max} \tau_{g}$ corresponding to ultra-slow pulses.
In experiments  involving
ultra-cold \cite{hau99} and hot \cite{lukin97} atoms  values of this
product on the order of few tens have been observed.
Potentially, an increase by two to three orders of magnitude  is likely,
which should allow for a high fidelity of entanglement.
We conclude by noting that exceptionally large nonlinearities have already
been measured in various experiments \cite{hau99,kash99,lukin97}. In
particular, an efficient nonlinear phenomenon
corresponding to  $\pi$ phase shift has
been observed with
pulse energies corresponding to less than $10^3$ photons per atomic cross
section
\cite{zibrov99}.

We thank Steve Harris for his encouragements and
acknowledge stimulating discussions with him, M.~Fleischhauer, R.~Glauber,
M.~O.~Scully,
M.~Werner and R. Walsworth.  This
work was supported by  the National Science Foundation, a David and Lucile
Packard Fellowship, and the US Army Research Office.

\def\etal{\textit{et al.}}


\begin{figure}[ht]
 \vspace*{2ex}
 \caption{
 	\label{f1.fig}
Prototype composite atomic medium for strong nonlinear interactions
at the single-photon level. By adjusting detunings such that the
two weak fields $E_{1},E_2$ and two classical fields
$\Omega_{1},\Omega_2$, are in resonance with transitions in
atoms A and B as shown,
resonant absorption can be eliminated and the group velocities of two weak
pulses can be made equal. The scheme can be implemented, for example,
using a natural  mixture of atomic rubidium. It consists of two isotopes
Rb$^{87}$ (natural abundance $\sim$ 25\%) and Rb$^{85}$ ($\sim$ 75\%) with
transition frequencies differing due to isotope shifts and nuclear spins. 
Propagation of tightly
focused beams over long distance can be achieved
using non-diffracting beams of Bessel shape,
fiber waveguides, or focusing properties associated with EIT
to induce waveguiding optically.
 }
\end{figure}


\begin{figure}[ht]
 \vspace*{2ex}
 \caption{
 	\label{f2.fig}
(a) Spectrum of transmission and refractive index corresponding to EIT.
Quantum interference in  the atomic ensemble induced by a
coherent driving field creates a sharp
resonance in the transmission of a weak optical field, accompanied by
rapid variation of the refractive index (red curve).
This rapid variation causes a dramatic reduction of group
velocity. The presence of a second weak field causes an effective shift of
the resonant frequency (green curve), which results in corresponding change of
a refractive index and hence the phase of the first weak field.
(b) Propagation dynamics in coherent media: when resonant pulses
enter the medium, they exhibit spatial compression and very soon
slip behind the reference pulse (white) that does not interact with the atoms.
The total time that two slow pulses can spend in the medium  is limited by
the residual single photon loss,
and by the spreading of the pulses
after $\tau_s \sim \Omega^2T^2/\gamma_{ab}$. The spreading is due to a
finite bandwidth $\Delta\omega_{max}$ of the EIT resonance.
 }
\end{figure}

\newpage

\begin{figure}[ht]
 \centerline{\epsfig{file=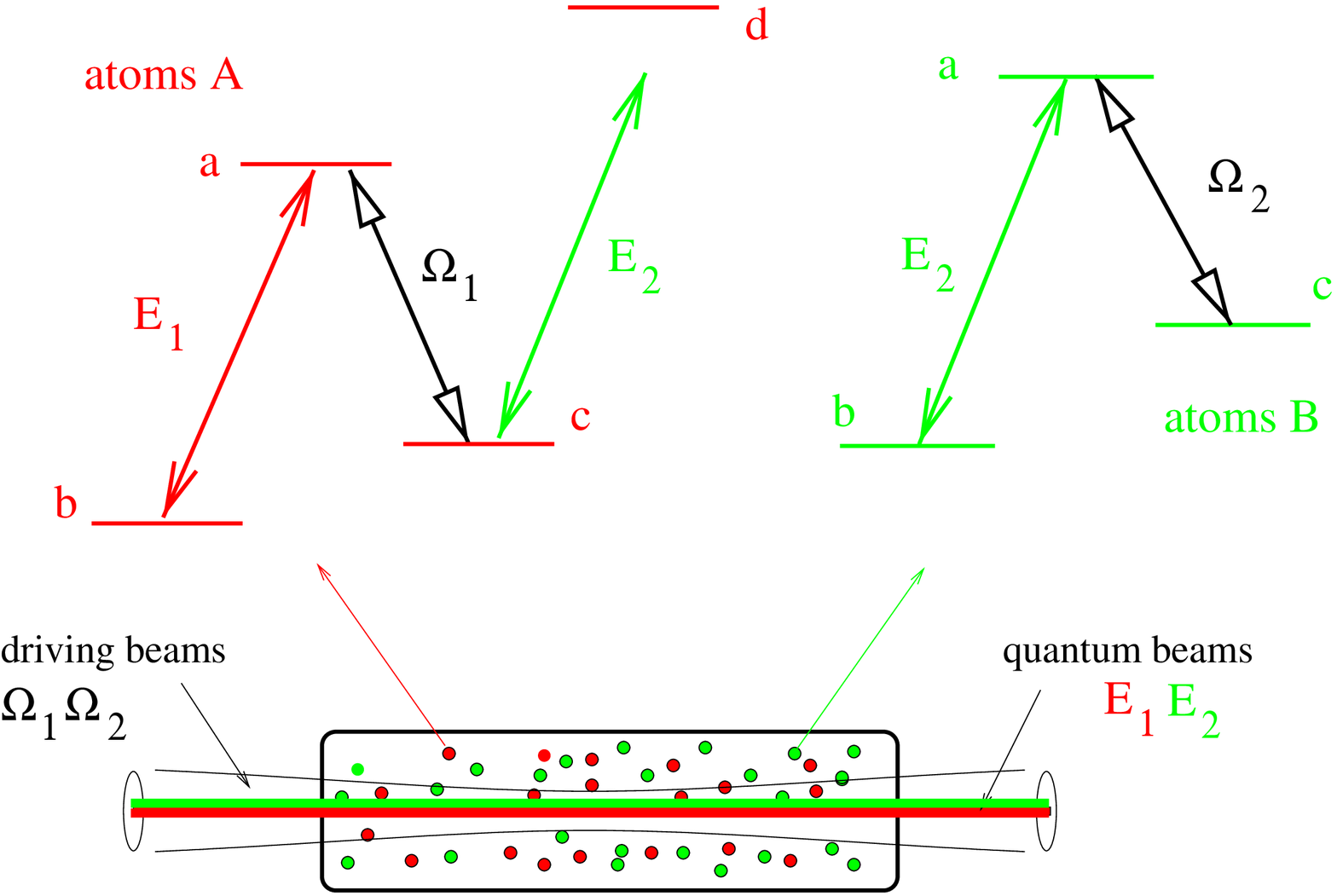,width=8.6cm}}
 \vspace*{2ex}
\end{figure}
\centerline{Figure 1}

\newpage

\begin{figure}[ht]
 \centerline{\epsfig{file=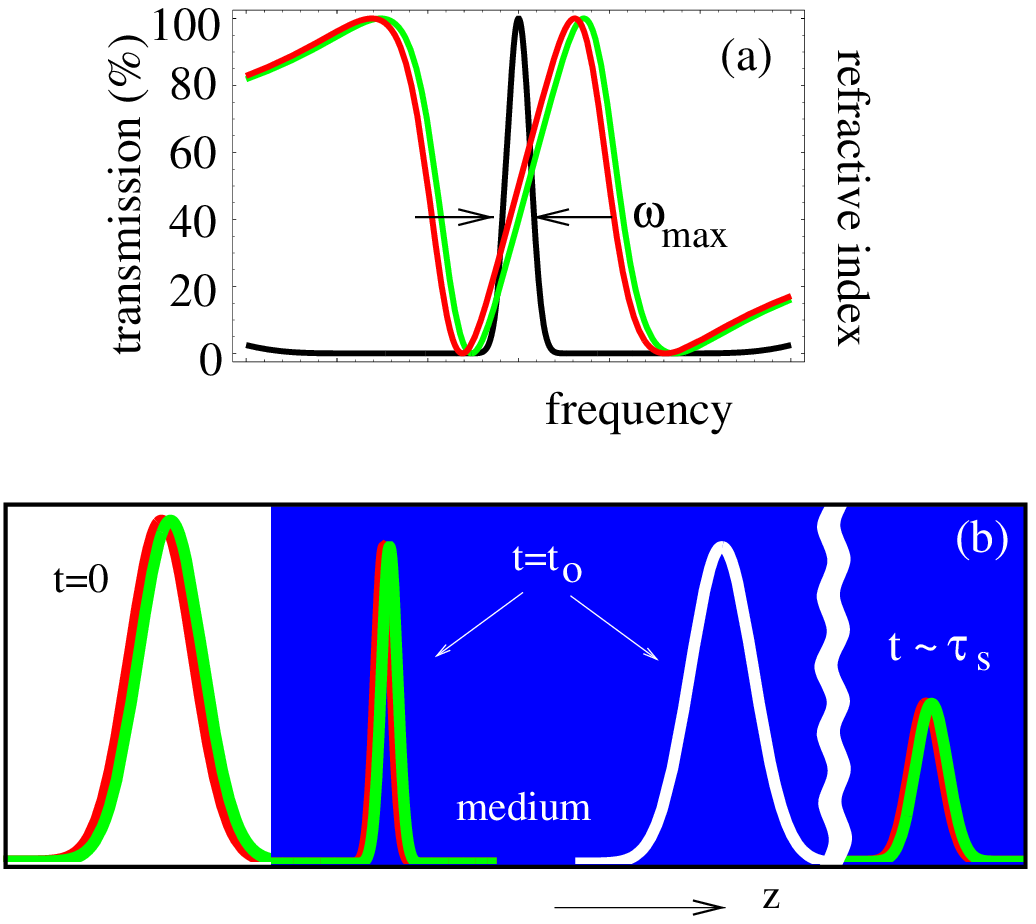,width=8.6cm}}
 \vspace*{2ex}
%
\end{figure}
\centerline{Figure 2}

\end{document}